# Programmable polyaniline nano neural network.
# A simple physical model.

Jerzy J. Langer

Laboratory for Materials Physicochemistry and Nanotechnology,
Faculty of Chemistry, Adam Mickiewicz University in Poznań,
Uniwersytetu Poznańskiego 8, 61-614 Poznań, Poland
Department of Physics of Functional Materials,
Faculty of Physics, Adam Mickiewicz University in Poznań,
Uniwersytetu Poznańskiego 2, Poznań 61-614, Poland
langer@amu.edu.pl

Abstract
Here we report a working physical model of a programmable neural nanonetwork based on a nanostructured conductive polymer - polyaniline. The device can be programmed owing to controlled growth of nanowires, a method developed in our laboratory. The response (output signal) of the network depends on the activated input line, but also on the amplitude of the input signals. The system is able to analyze multiple input signals in parallel and in real time. This is still considered as a challenge.

Artificial intelligence (AI) is currently a hot topic among scientific, technical and technological issues. Hardware requirements, in the light of the rapidly growing interest and applications of AI, are very high, challenging and often even bordering on unfeasible.
In addition, the growing necessary energy consumption is a serious problem.
The expected solution is parallel information processing and deep miniaturization at the nano scale, including the molecular level. The goal of our project is to push the boundaries in the construction of physical neural networks as miniature devices based on conductive nanofibers created randomly, but in a deliberately organized network. In particular, we focus on the design, creation and testing of experimental models of polyaniline neural nano networks [11, 12], also improving the controlled growth of nanofibers to achieve optimal stability, repeatability and adaptability. To obtain the highest quality nanofibers, the most effective and promising are experiments conducted in reduced gravity, in our case using a drop tower [15].
Our approach, based on the properties of a network of polyaniline nanofibers, created in a controlled way, making the device programmable, is a step towards progress in most of the considered aspects.

Polyaniline Nano Neural Network (P3N)
It is a miniature logic device that can analyze parallel input signals in real time. Despite a large number of published works on nano wiring, nano networking [1-4, 7-10] and nano neural networks [5, 6], the described system is one of the first working models of a physical nano neural network (Fig. 1, 2), being developed in our Laboratory step by step since 2000. The results of preliminary preparatory experiments were published and presented at conferences on an ongoing basis [7-12,13-15].

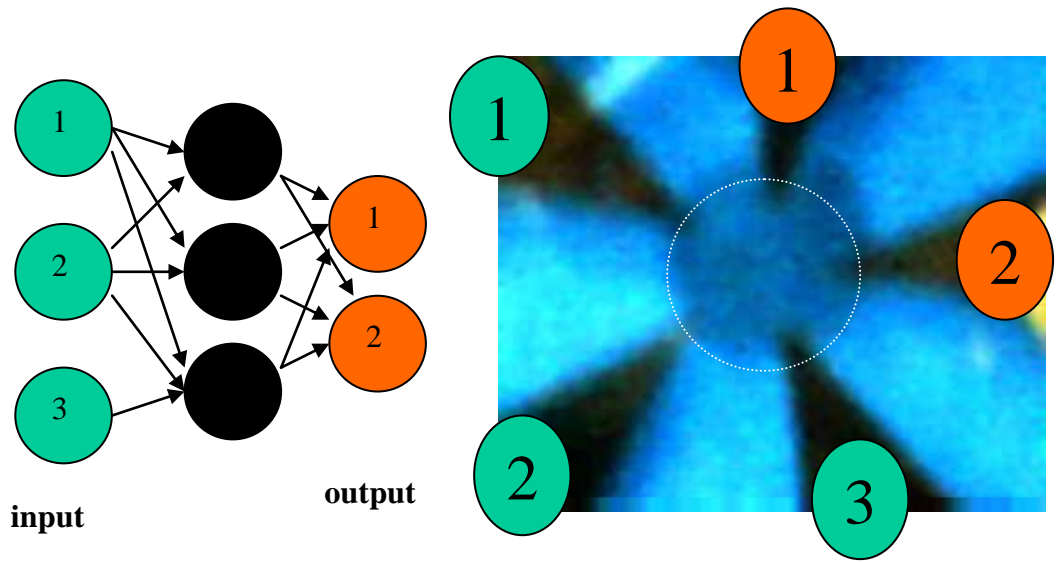

**Fig. 1**. Schematic (left) and physical model of the experimental P3N polyaniline neural nanonetwork (right) viewed under a transmission optical microscope (50x).

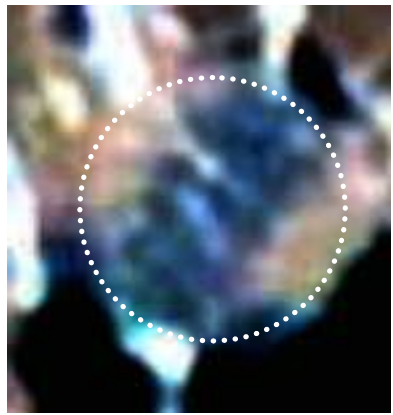

**Fig. 2**. Active area of the P3N neural network after the learning process (image from an optical microscope, 50x). A deliberately organized network of polyaniline nanofibers is clearly visible (blue Y-shaped area).

The experimental polyaniline nano neural network (P3N) is based on a self-assembled random network of nanofibers that is intentionally formed (organized) in a controlled manner (Fig. 3 and 4) - as opposed to the solvent-mediated approach, e.g. Gruner and Nugent [5,6]. Our experimental thin-film (quasi-2D) logic chip consists of 3 input thin-film gold microelectrodes, 2 output thin-film gold microelectrodes and a network of polyaniline nanofibers in between (of 1-3 mm working area; Fig. 1-2). The device can be programmed using electrically driven, controlled nanowire growth (Fig. 2-3), a method developed in our laboratory [7-12]. The best, highest quality nanofibers can be obtained in reduced gravity (Fig. 4b), in our case using a 44 m high drop tower (falling time 3 s) [15].

The results of the research experiments are summarized in Table 1 and Table 2. The response (output signal) of the P3N device depends on the activated input line, but also on the amplitude of the input signals. Thanks to this, the system is able to analyze input signals in parallel, in a way similar to that observed in a natural neural network (including the most complex one - the brain) and in real time.
Generally, the answer is not additive - typically, the measured values are less than the arithmetic sum of the isolated corresponding contributions. However, in two cases we observe results very close to additivity: at output 2 with a 3 V input signal supplied simultaneously by inputs 1 and 3 (0.49 V vs 0.48 V) or inputs 2 and 3 (0.38 V vs 0.35 V), Table 1.
In the working mode, the device effectively operates in a voltage regime consistent with traditional electronic devices (e.g. 5 V), but also at very low voltage (mV), which is biocompatible. This is a very promising feature. The device recognizes low input signals of various amplitudes, delivered in various ways - in various combinations and through various input channels (Table 2).
The learning mode usually requires the use of a different voltage range, depending on the properties and functionality of the specific nanonetwork, in particular its functional flexibility, which needs the reversible processes to be generated.

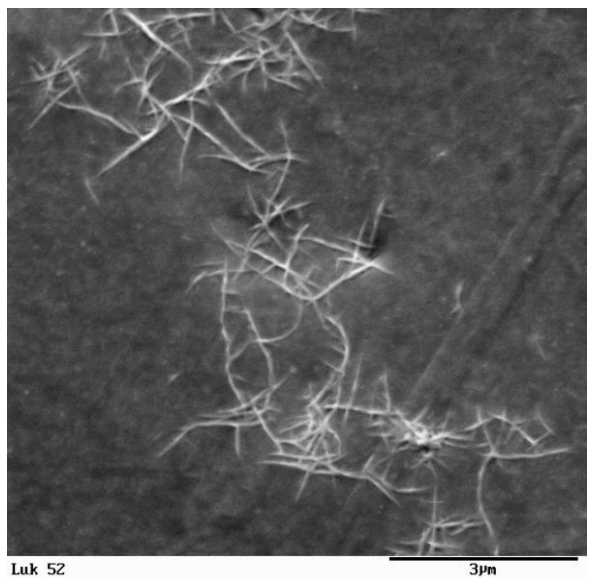
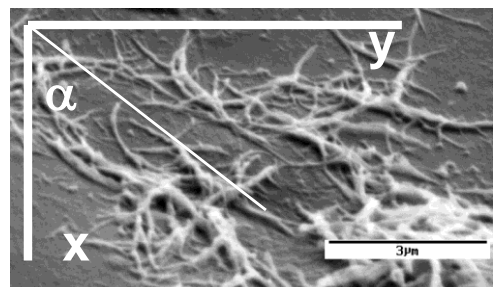
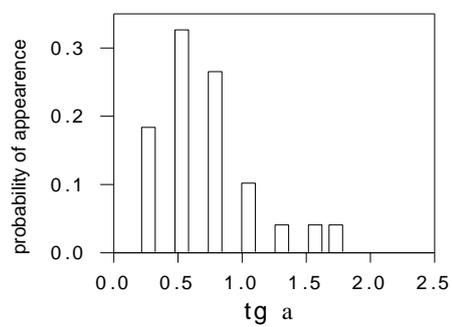

**Fig. 3**. Nanofibrils intentionally formed and oriented in a conductive microchannel: SEM micrographs and statistical analysis.

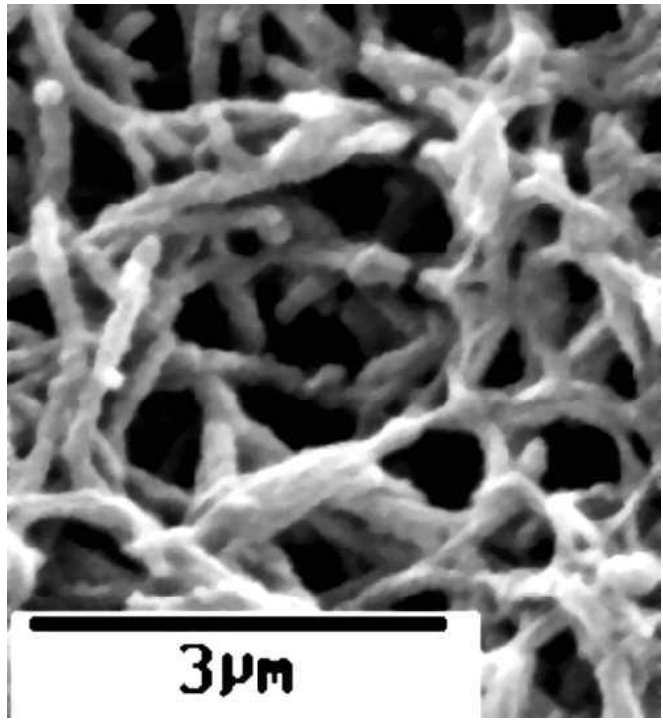

**(a)**

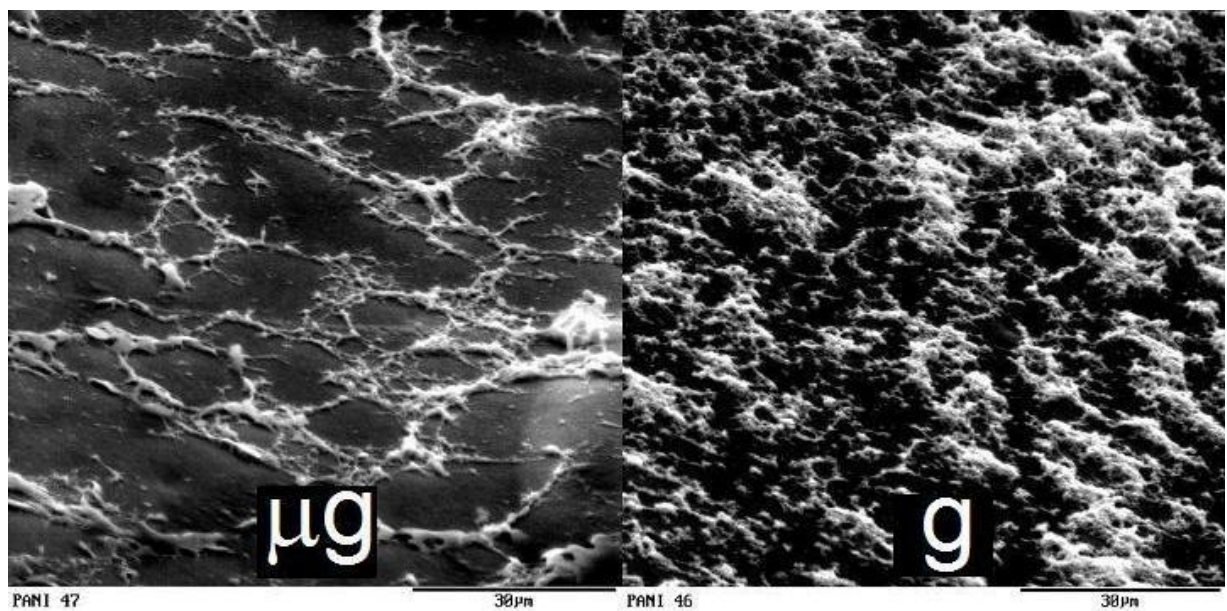

**(b)**

**Fig. 4**. Densely packed network of polyaniline nanofibers (a); influence of microgravity [15] (b) – left. SEM micrographs.

**Table 1**. Electrical response ([V]) of the P3N system tested for a 3 V and 4 V input signals.

|       | Input signal amplitude |       |       |       |
|-------|-------|-------|-------|-------|
|       | 3V    |       | 4V    |       |
|       | Output [V] | | Output [V] | |
|       | 1     | 2     | 1     | 2     |
| **1** | 1,03  | 0,37  | 1,39  | 0,52  |
| **1+2** | 1,08<br>arithmetic 1,56 | 0,41<br>arithmetic 0,61 | 1,45<br>arithmetic 2,11 | 0,56<br>arithmetic 0,95 |
| **1+3** | 1,11<br>arithmetic 1,26 | **0,49**<br>arithmetic **0,48** | 1,51<br>arithmetic 1,79 | 0,68<br>arithmetic 0,82 |
| **2** | 0,53  | 0,24  | 0,72  | 0,43  |
| **2+3** | 0,65<br>arithmetic 0,75 | **0,38**<br>arithmetic **0,35** | 0,90<br>arithmetic 1,12 | 0,53<br>arithmetic 0,73 |
| **3** | 0,23  | 0,11  | 0,40  | 0,30  |

**Table 2**. P3N response pattern observed with a very low voltage applied (input signal: 5-50 mV).

| | Input amplitude | | | | | | | |
|---|---|---|---|---|---|---|---|---|
| | 5 mV | | 10 mV | | 30 mV | | 50 mV | |
| | Output | | Output | | Output | | Output | |
| Input | 1 | 2 | 1 | 2 | 1 | 2 | 1 | 2 |
| 1 |   | ■ |   | ■ |   | ■ |   |   |
| 2 |   |   |   | ■ | ■ |   |   |   |
| 3 | ■ |   | ■ | ■ |   | ■ | ■ | ■ |

■ ≥ 1 mV,   □ ~0 mV

The main part of the device is a network of polyaniline micro- and nanofibers that remain in good electrical contact with gold microelectrodes, and the whole is formed on a plastic or glass substrate (for mechanical protection).
In the first stage, thin-film gold microelectrodes are deposited on a substrate (plastic foil or glass plate) by evaporating metallic gold in a vacuum using a mask with an appropriate pattern. Then, a nanostructured polyaniline layer is electrochemically prepared on gold electrodes and in between at a potential of 1V (vs. Ag/AgCl, potentiostat EP20A, ELPAN or Autolab, Metrohm Autolab). The electrolyte is a 10% aqueous solution of aniline hydrochloride at pH ~1 (HCl). The thickness of PANI network layer, the number of nanofibers and their packing density, are controlled by a limited charge flow (the reaction time is controlled and set in the range of 0-300 s). Freshly prepared polyaniline micro- and nanofibril network is washed with distilled water and then dried on air.
This is a standard procedure applied in our laboratory, including subsequent characterization of the material formed (polyaniline), as described elsewhere [13,14].
Then the network is modified in the learning procedure using low-voltage pulses (1-5 V). This leads to local changes in the electrical properties of the nanofibers depending on the required function.
Alternatively, wet learning is performed during the preparation of the nanofiber networks, based on preliminary experiments with controlled nanofiber formation (Fig. 3) [7-13].

Microgravity experiments were carried out using a drop tower with a fall time of 3 s and twin automatic electrochemical reactors (one - dropped - equipped with a gravity sensor and a radio link enabling turning on and off the process in both reactors), operating under identical standard conditions, except for gravity [15].

Conclusion

In concept, we are close to a natural solution. The network of polyaniline nanofibers (Fig. 4), each of which has a different electrical conductivity, corresponds to a network of synaptic connections of different properties and importance (weight). While, the nonlinearity of electrical contacts between polymer nanofibrils mimics the transfer function of neurons [16].

The learning process transforms the P3N structure: the initial almost symmetrical shape (Fig. 1) can change into an asymmetric one after learning (e.g. similar to Y-like shape, Fig. 2). This may be reversible by an electrically induced oxidation-reduction reaction in polyaniline nanofibers. The oxidized form of PANI is conductive and dark blue, while the reduced form is almost colorless and non-conductive. This makes P3N programmable and flexible in terms of functionality.

This is the first step, a simple but promising logical microdevice with possible applications in the field of artificial intelligence (AI). The advantage is direct parallel processing of information, but also real-time operation with both continuous and discrete input signals.
The success of experiments with P3N shows that physical polyaniline nano neural networks (generally - polymer nano neural networks) provide the opportunity to build new, programmable systems with parallel information processing and significant potential in AI applications.

Acknowledgments
In memory of Anna, an unforgettable beloved wife, for her priceless, unlimited support.

We would like to thank students Łukasz Markowski, Michał Wassel, Łukasz Wolny, Jacek Włodarczak and Magdalena Michalewicz, who partially collaborated on this project during their master's theses in our laboratory.